# Unusual Exciton-Phonon Interactions at van der Waals Engineered Interfaces

*Colin M. Chow[1], Hongyi Yu[2], Aaron M. Jones[1], Jiaqiang Yan[3], David G. Mandrus[3,4], Takashi Taniguchi[5], Kenji Watanabe[5], Wang Yao[2]\*, Xiaodong Xu[1,6]\**

[1]Department of Physics, University of Washington, Seattle, Washington 98195, USA.

[2]Department of Physics and Centre of Theoretical and Computational Physics, University of Hong Kong, Hong Kong, China.

[3]Materials Science and Technology Division, Oak Ridge National Laboratory, Oak Ridge, Tennessee 37831, USA.

[4]Department of Materials Science and Engineering, University of Tennessee, Knoxville, Tennessee 37996, USA.

[5]Advanced Materials Laboratory, National Institute for Materials Science, Tsukuba, Ibaraki 305-0044, Japan.

[6]Department of Materials Science and Engineering, University of Washington, Seattle, Washington 98195, USA.

**Abstract**

Raman scattering is a ubiquitous phenomenon in light-matter interactions which reveals a material's electronic, structural and thermal properties. Controlling this process would enable new ways of studying and manipulating fundamental material properties. Here, we report a novel Raman scattering process at the interface between different van der Waals (vdW) materials as well as between a monolayer semiconductor and 3D crystalline substrates. We find that interfacing a $WSe_2$ monolayer with materials such as $SiO_2$, sapphire, and hexagonal boron nitride (hBN) enables Raman transitions with phonons which are either traditionally inactive or weak. This Raman scattering can be amplified by nearly two orders of magnitude when a foreign phonon mode is resonantly coupled to the A exciton in $WSe_2$ directly, or via an $A'_1$ optical phonon from $WSe_2$. We further showed that the interfacial Raman scattering is distinct between hBN-encapsulated and hBN-sandwiched $WSe_2$ sample geometries. This cross-platform electron-phonon coupling, as well as the sensitivity of 2D excitons to their phononic environments, will prove important in the understanding and engineering of optoelectronic devices based on vdW heterostructures.

Two-dimensional (2D) quantum materials, with their wide range of physical properties, have emerged as an exciting platform for exploring new science and technologies. Being atomically-thin, their material properties and performance in devices are inevitably affected by the environment. For instance, the mobility of $SiO_2$-supported graphene is ultimately limited by the polar phonon in $SiO_2$[1], while hBN provides an atomically-smooth substrate which enables ultra-high mobility graphene devices[2]. The van der Waals (vdW) nature of such 2D materials enables interface engineering, a powerful approach to probe emerging phenomena. Outstanding examples include the observation of high $T_c$ superconductivity in monolayer FeSe on STO substrates[3], the demonstration of Hofstadter's butterfly physics in crystallographically aligned graphene-hBN superlattices[4], and the observation of valley dependent many-body effects and ultra-long valley polarization lifetimes[5,6] in 2D semiconductor heterostructures. While efforts have been primarily focused on understanding and developing electronic, photonic, and spin-valley properties, the effects of microscopic vibrational modes (phonons) in 2D materials, in particular, on their optoelectronic properties, are much less understood. Due to a reduced screening effect in atomically-thin materials, one would expect that the interaction of excitons with its phononic environment (such as optical phonons in the substrates) would become significant through Fröhlich (or dipole-dipole) interactions, potentially leading to new phenomena unobserved in their constituent compounds.

The study of phonons in solids typically employs Raman or infrared (IR) spectroscopy. In general, due to symmetry restrictions, only a subset of all possible phonon modes are Raman or IR active. For instance, the $A_{2u}$ mode in hBN has odd parity under inversion[7] and creates an out-of-plane dipole, which is IR active but Raman silent[7,8]. In contrast, the $B_{1g}$ mode, which has even parity under inversion, is both IR and Raman silent. Unlike IR absorption, Raman scattering involves inelastic photon scattering and is typically weak. Nonetheless, it can be enhanced in many ways, for example, by matching the energy of either the incident or inelastically scattered photon with an electronic transition. Such resonantly enhanced Raman scattering has been reported in graphene[9,10] and monolayer semiconductors[11,12] for both Stokes and anti-Stokes[13,14] processes, where the phonon coupling mechanism primarily occurs within the same material.

In this work, we report a novel Raman scattering phenomenon at the vdW interface formed between monolayer $WSe_2$ and three popular materials for building 2D optoelectronic devices: atomically-thin hBN, $SiO_2$, and sapphire. To fabricate the samples, we mechanically exfoliate monolayer $WSe_2$ and atomically thin (~ 10 layers) hBN onto 300-nm-thick $SiO_2$ on Si substrates, and then stack them using dry transfer techniques. Five sample configurations are studied: (i) $WSe_2$ on a bare $SiO_2$ substrate ($WSe_2/SiO_2$), (ii) hBN-covered $WSe_2$ ($hBN/WSe_2/SiO_2$), (iii) hBN-sandwiched $WSe_2$ ($hBN/WSe_2/hBN/SiO_2$), (iv) $WSe_2$ on a bare sapphire substrate ($WSe_2$/sapphire), and (v) $WSe_2$ on an hBN underlay ($WSe_2/hBN/SiO_2$). In the main text, we present experimental results from the first four structures, some of which are illustrated in Figure 1a. The result from the $WSe_2/hBN/SiO_2$ configuration is qualitatively similar to that of $hBN/WSe_2/SiO_2$ and is presented in the supplementary information.

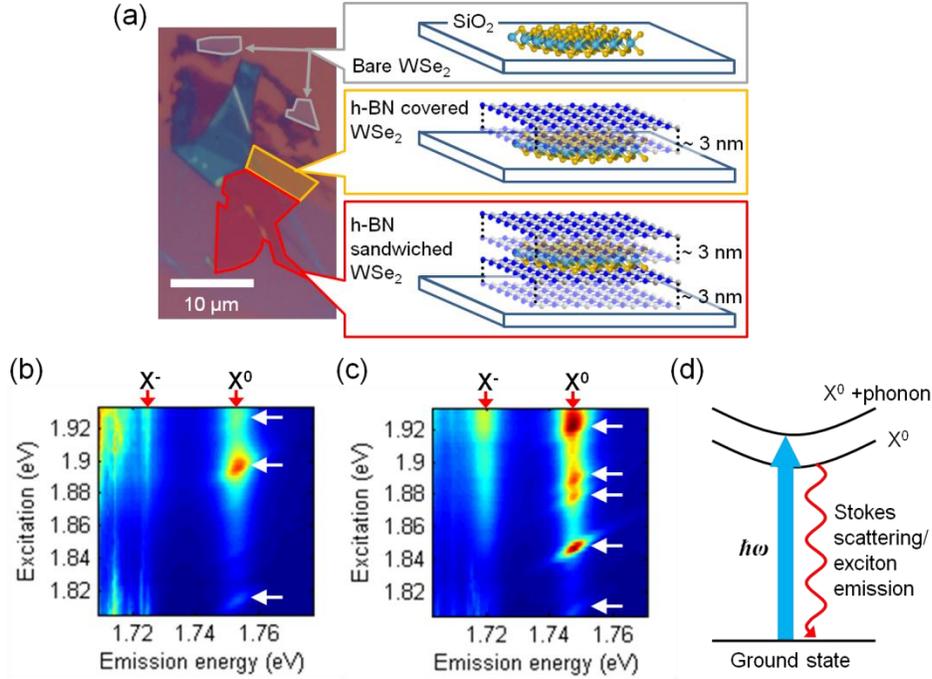

**Figure 1.** (a) Optical micrograph of the sample with three configurations highlighted: gray for $WSe_2/SiO_2$, yellow for $hBN/WSe_2/SiO_2$, and red for $hBN/WSe_2/hBN/SiO_2$. (b, c) PLE intensity plots from $WSe_2/SiO_2$ and $hBN/WSe_2/SiO_2$ samples, respectively. Neutral exciton ($X^0$) and trion ($X^-$) are marked. PLE peaks are indicated by the white arrows in $X^0$ emission, with two new resonances emerging in the $hBN/WSe_2/SiO_2$ structure. (d) Energy level diagram illustrating resonant Raman scattering via coupling to the neutral exciton in $WSe_2$.

We perform photoluminescence (PL) measurements on all sample configurations while varying the laser excitation energy, i.e. PLE spectroscopy, at a temperature of 5K. Figure 1b,c show PLE intensity plots from $WSe_2/SiO_2$ and $hBN/WSe_2/SiO_2$ samples, respectively, in which the $X^0$ and trion ($X^-$) emission peaks are indicated. In both cases, the response of $X^0$ luminescence to the excitation laser energy features pronounced resonances, while that of $X^-$ is monotonic[15]. As Figure 1b shows, there are three prominent peaks in the $X^0$ PLE of the $WSe_2/SiO_2$ sample, as indicated by white arrows. In contrast, the $X^0$ PLE becomes remarkably different in $hBN/WSe_2/SiO_2$ structure, where two additional peaks emerge. Nonetheless, all resonances shift in parallel with the excitation laser energy, implying their origin in Raman-scattering. Note that these features are also present on both sides of the $X^0$ resonance (see supplementary information), albeit with much weaker intensities, and shift with the excitation laser as with ordinary Raman scattering. The enhancement seen at the $X^0$ resonance is due to resonant Raman scattering, where the Stokes-shifted emission energetically matches the $X^0$ resonance, as illustrated in Figure 1d.

To understand the physical origin of these resonant Raman peaks, we recast the data of Figure 1b,c in terms of excess photon energy (see Figure 2a,b), which is defined as the energy difference between the excitation source and emission. In this way, we can remove the effects of small shifts in $X^0$ peak position which stem from varying dielectric environments between sample geometries. Consistent results from a second set of samples are presented in the supplementary information. A comparison between Figure 2a,b shows that there are three common resonances near excess energies of 173, 141, and 62 meV, with ±1 meV uncertainty. Additionally, in hBN/WSe$_2$/SiO$_2$ heterostructures, two PLE peaks appear at excess energies of 130 and 98 meV. Figure 2c,d show vertical line cuts at the $X^0$ resonances where relative intensities and linewidths of the PLE peaks are apparent. Among them, the linewidths of 98 and 62 meV peaks are found to be about 5.5 meV (limited by the 5.5 meV laser linewidth), significantly narrower than the others (~ 15 meV). This implies that these two peaks solely arise from resonant Raman scattering involving the $X^0$ transition. Were they to originate from excitonic states at corresponding excess energies, one would expect much broader spectral widths, on the order of 10 meV[16].

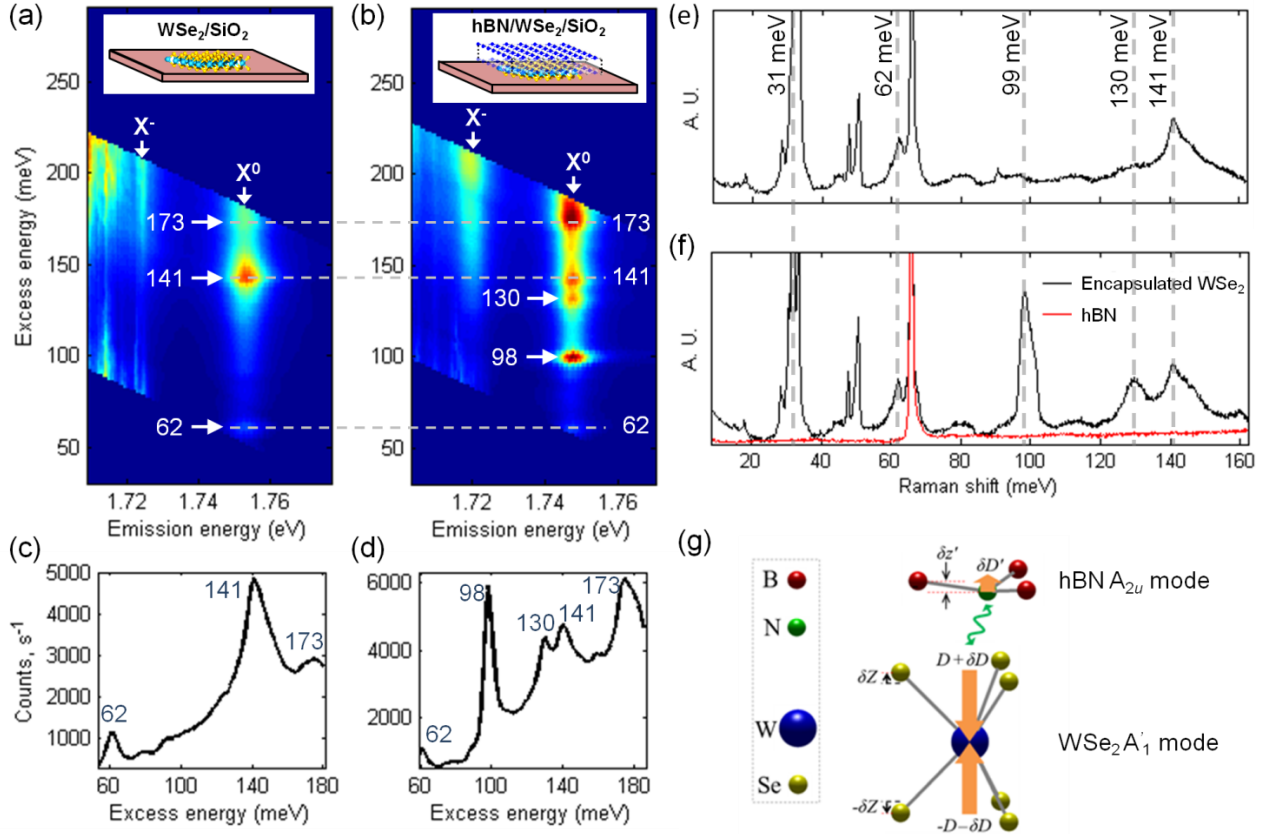

**Figure 2.** (a, b) PLE intensity plots of WSe$_2$/SiO$_2$ and hBN/WSe$_2$/SiO$_2$ samples, respectively, plotted with the excess energy on the vertical axis. The excess energies for prominent PLE peaks are indicated next to white arrows. Horizontal dashed lines indicate common PLE peaks seen in both sample geometries. Insets: schematics of sample geometries. (c, d) Vertical line cuts at the $X^0$ resonances from the PLE maps in (a) and (b), respectively, with the energies of PLE peaks

indicated. (e) Raman spectrum from WSe$_2$/SiO$_2$ sample. See supplementary information for mode assignments. (f) Raman spectra from hBN/WSe$_2$/SiO$_2$ sample (black) and an adjacent atomically-thin hBN (red). (g) Schematic of the dipole-dipole interaction model. The hBN A$_{2u}$ phonon corresponds to relative displacements $\delta z'$ between B and N atoms, resulting in an electric dipole $\delta D'$. Here, the WSe$_2$ atomic configuration can be modeled as two static electric dipoles $\pm D$. The wavy arrow represents the interaction between $\delta D'$ and $D$. The WSe$_2$ A$_1'$ phonon mode imparts a small change, $\delta D$, to the magnitudes of $\pm D$.

To identify the phonon modes observed in the PLE spectra, we performed Raman spectroscopy using 532-nm excitation. The results are shown in Figure 2e,f for WSe$_2$/SiO$_2$ and hBN/WSe$_2$/SiO$_2$ geometries, respectively. Both samples prominently feature the A$_1'$ mode at 31 meV in WSe$_2$, along with its overtone (2$^{nd}$ harmonics) at 62 meV, which is responsible for the PLE peaks at the corresponding excess energy. The phonon mode at 141 meV in both spectra will be addressed later. Remarkably, two new Raman signals emerge in the hBN/WSe$_2$/SiO$_2$ configuration at 130 and 98 meV, energetically matching the corresponding peaks in PLE. While there is no experimental indication of a phonon mode in WSe$_2$ at these energies, the IR active A$_{2u}$ ZO mode in hBN has been observed at 97.1 meV[17], which is energetically close to the 98 meV resonances seen in Figure 2b,d. Besides, the Raman and IR silent B$_{1g}$ ZO mode is also expected to be energetically close to 98 meV[8]. For comparison, the Raman spectrum from atomically-thin hBN adjacent to the hBN/WSe$_2$/SiO$_2$ sample is also shown in Figure 2f, in which the signal at 98 meV is absent. Thus, our results indicate that Raman silent hBN phonon modes are activated in hBN/WSe$_2$/SiO$_2$ vdW heterostructures, where in PLE measurement the signal is enhanced by nearly two orders of magnitude through interfacial coupling to X$^0$ in WSe$_2$. (Supplementary information) The activation of Raman scattering is due to reduced symmetry from the interaction between hBN and WSe$_2$. Regarding the origin of the 130-meV resonance, we note that although neither hBN nor WSe$_2$ has a phonon mode at this energy, it is approximately given by the combined energy of hBN A$_{2u}$ and WSe$_2$ A$_1'$. We therefore posit that the resonance at 130 meV is the result of a three-way interfacial coupling between A$_{2u}$ (or B$_{1g}$) in hBN, and A$_1'$ phonon and X$^0$ in WSe$_2$.

We further explore the effect of interfacial exciton-phonon coupling in hBN-sandwiched WSe$_2$ samples, i.e. the hBN/WSe$_2$/hBN/SiO$_2$ geometry illustrated in Figure 1a. Figure 3a shows the PLE intensity map where the vertical coordinate is the excess energy, as in Figure 2a,b. Again, this makes comparison with preceding PLE data sensible by removing the effects of energy shifts in the X$^0$ resonance, which in this case has become significant, as Figure 3b shows. The vertical line cut in Figure 3c shows the PLE peaks at the X$^0$ resonance. Compared to the hBN/WSe$_2$/SiO$_2$ sample, a remarkable difference is that the signal from the hBN phonon resonance at 99 meV is suppressed, while that of hBN A$_{2u}$ + WSe$_2$ A$_1'$ at 131 meV is enhanced (evident in Figure 3b). Consistent with the PLE data, the Raman spectrum given in Figure 3d also shows a suppressed 98 meV mode and a pronounced signal at 131 meV. However, no clear indication of resonance exists at 141 meV. The stark difference of both Raman and PLE spectra between

hBN/WSe$_2$/SiO$_2$ and hBN/WSe$_2$/hBN/SiO$_2$ samples demonstrates the dependence of interfacial exciton-phonon coupling in vdW heterostructures on sample environment, and, consequently, its manipulation by vdW engineering. Notably, a study in hBN-sandwiched WSe$_2$ heterostructures is recently reported[18], where interfacial exciton-phonon scattering is also identified. The report shows a weaker 98-meV mode relative to the 131-meV peak, which agrees with our observation. Here, by examining different sample geometries, we are able to reveal that this is caused by the suppression of the interaction between hBN ZO phonons and WSe$_2$ exciton due to symmetry reconstruction in the hBN sandwiched WSe$_2$ sample, as discussed below.

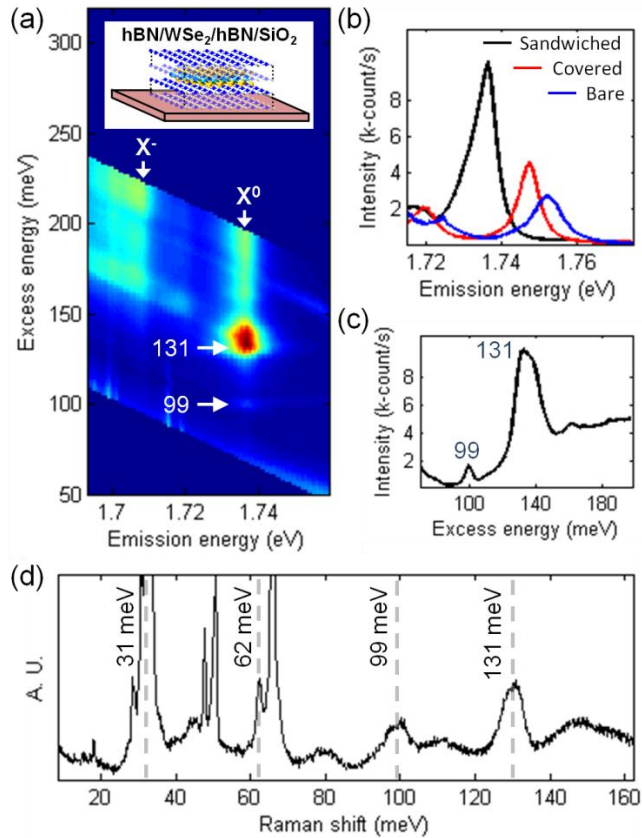

**Figure 3.** (a) PLE intensity plot of hBN/WSe$_2$/hBN/SiO$_2$ as a function excess and emission energies. Inset: schematic of sample geometry. (b) Emission spectra of hBN/WSe$_2$/hBN/SiO$_2$ (black), hBN/WSe$_2$/SiO$_2$ (red) and WSe$_2$/SiO$_2$ (blue) samples at the excess energy of 130 meV, showing the energy shifts of the X$^0$ resonance for different sample configurations. (c) Vertical line cut at the X$^0$ resonance from the PLE map in (a). (d) Raman spectrum from the hBN/WSe$_2$/hBN/SiO$_2$ sample.

While a microscopic theory for the aforementioned observations has yet to be developed, we suggest here a possible mechanism, where the interlayer coupling between hBN phonons and WSe$_2$ excitons is mediated by WSe$_2$ out-of-plane phonons, in particular, the $A'_1$ mode and the flexural (ZA) mode. Here, the hBN $A_{2u}$ or $B_{1g}$ phonon is associated with an out-of-plane electric dipole $\delta D'$, which interacts with the monolayer $WSe_2$, whose charge distribution in the tri-atomic planes resembles two layers of dipoles in opposite orientations, $\pm D$, that are changed by the out-of-plane $A'_1$ (or ZA) phonon mode (Figure 2g). The dipole-dipole interaction can then couple the hBN and $WSe_2$ phonons, which then interact with the exciton in $WSe_2$. Our analysis finds that the $WSe_2$ ZA mode can play a key role in the generation of the observed 98 meV peak, where the difference in the PLE spectra between the hBN/$WSe_2$/$SiO_2$ and hBN/$WSe_2$/hBN/$SiO_2$ geometries can be explained by considering the out-of-plane mirror ($\hat{\sigma}_h$) symmetry of the $WSe_2$ samples. In the hBN/$WSe_2$/hBN/$SiO_2$ sample geometry, the $\hat{\sigma}_h$ symmetry is present (or weakly broken), which suppresses the interaction between the exciton and the ZA mode (note that the ZA mode is antisymmetric with respect to $\hat{\sigma}_h$, while the exciton is symmetric). This leads to a reduction of the 98 meV signal. Conversely, the 130 meV signal, which largely arises from the $WSe_2$ $A'_1$ mode, is symmetric under $\hat{\sigma}_h$. Hence a strong 130 meV signal is observed in both samples.

Finally, we refer back to the PLE intensity map of Figure 2a, where the origin for enhanced Raman peaks at 141 and 173 meV remains to be determined. The resonance feature at 141 meV has attracted attention recently due to its apparent role in the generation of valley coherence in monolayer $WSe_2$[19]. Nonetheless, its physical origin remains unclear. On one hand, it has been ascribed to the nearly degenerate 2s/2p excitonic states that energetically coincide with a combination of multiple phonons in $WSe_2$[19,20]. On the other hand, studies have suggested that the 2s/2p states have a significant energy difference and lie at least 160 meV above the $X^0$ resonance[16,21]. Coincidentally, thermally grown $SiO_2$ has been found to show a pronounced surface phonon mode at about 141 meV[22,23]. This strongly implies that the 141 meV in our measurements originates from the $SiO_2$ surface phonon, which resonantly couples to $X^0$ in $WSe_2$, and furnishes the 141 meV peak in PLE. If correct, then the resonance at 173 meV simply arises from the combination of $WSe_2$ $A'_1$ and the $SiO_2$ surface phonon, consistent with the origin of similar PLE peaks in hBN/$WSe_2$/$SiO_2$ samples. Our PLE data from hBN/$WSe_2$/hBN/$SiO_2$ (Figure 3c) and $WSe_2$/hBN/$SiO_2$ (supplementary information) samples support such an interpretation. They show that the PLE intensity at 141 meV is reduced relative to that at 130 meV. This is due to the insertion of hBN between $WSe_2$ and the $SiO_2$ substrate, which partially screens the excitonic coupling to substrate phonons.

To verify, we repeat our measurements on a $WSe_2$ monolayer placed on a sapphire substrate. With the exception of the 62-meV peak arising from the 2 $A'_1$ mode in $WSe_2$, the PLE results shown in Figure 4a,b reveal resonances distinct from all preceding cases. Notably, the 141-meV peak seen in previous samples is now absent. This unambiguously pinpoints the physical origin of the 141-meV peak to the $SiO_2$ surface phonon. Furthermore, two new resonances appear at 124 and 93 meV. By performing Raman spectroscopy on both the sapphire substrate (Figure 4c) and the $WSe_2$ monolayer (Figure 4d), the origin of these peaks can be traced to the $E_g$ optical phonon in sapphire[24]. Here, 93 meV matches the Raman active $E_g$ mode[25], while 124 meV is simply the combination of the $E_g$ mode in sapphire and $A'_1$ mode in WSe$_2$.

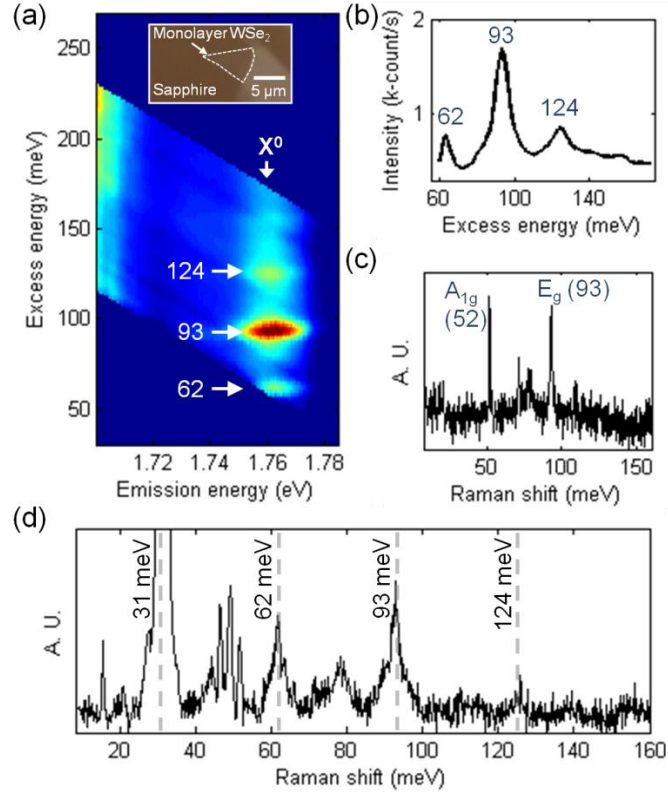

**Figure 4.** (a) PLE intensity plot of monolayer WSe$_2$ on sapphire substrate as a function of excess and emission energies. Inset: optical micrograph of WSe$_2$ on sapphire substrate, with the monolayer region indicated. (b) Vertical line cut at the $X^0$ resonance from the PLE plot in (a). (c, d) Raman spectra from the sapphire substrate and the WSe$_2$/sapphire geometry, respectively.

In summary, we observe a novel resonant Raman scattering phenomenon via cross-material exciton-phonon coupling at vdW interfaces, in contrast to conventional resonant Raman scattering where the mechanism occurs within a given material. The sensitivity of excitons in monolayer materials to their phononic environments, such as those provided by SiO$_2$, hBN, and sapphire, can be utilized to enhance our understanding of atomically-thin devices. Moreover, this also reveals that, given the close physical contact between 2D materials and their supporting substrates, the effects of substrate phonons can be significant. This largely ignored aspect of device construction may add to the challenges of 2D-material studies in which substrate contributions need to be isolated. Regardless, our work reveals the potential of vdW heterostructures as a tool for studying and designing Raman scattering in atomically-thin devices, as well as a platform for novel opto-phononic applications.

## Methods

Monolayer WSe$_2$ samples are mechanically exfoliated from bulk WSe$_2$ crystals onto 300-nm-think SiO$_2$ substrates thermally grown on doped Si wafers. The monolayers are first identified visually by their optical contrast under a microscope and then confirmed by atomic force microscopy. Similar exfoliation techniques are used to obtain atomically-thin hBN layers, where their thicknesses are chosen to be around 3 nm. Fabrication of the vdWHs is achieved with dry transfer techniques. As an example, to create the hBN/WSe$_2$/hBN/SiO$_2$ structure, first, the top hBN layer is picked up with a PPC film stretched over a dome-shaped glass epoxy. Next, using a micromanipulator system positioned under an optical microscope, the monolayer WSe$_2$ is aligned with the top hBN and picked up with the same PPC film. Finally, the hBN/WSe$_2$ stack is deposited onto the bottom hBN layer already exfoliated onto a SiO$_2$ substrate, via the hot-release method at 90 $^o$C.

During optical measurements, the samples are kept at a temperature of 5 K in a cold-finger cryostat. All optical studies are made in reflection geometry, where the photoluminescence is collected via a microscope objective mounted on a micrometer stage assembly. A supercontinuum fiber laser serves as the light source for PLE measurements. Wavelength selection is made possible with a tunable filter, which produces frequency-narrowed output pulses with a bandwidth of about 2 nm. Additional filters are placed in the beam path to remove the side lobes. The excitation beam is reflected onto the sample using a dichroic mirror, which also deflects the reflected excitation beam to prevent it from entering the spectrometer. A similar optical setup is used for Raman spectroscopy with 532-nm excitation, except that the supercontinuum laser is replaced by a CW laser and a high-contrast Bragg reflector is used in instead of the dichroic mirror. Both PLE and Raman spectra are recorded with a TE cooled CCD camera attached to the output port of the spectrometer.

ASSOCIATED CONTENT

**Supporting Information**. The Supporting Information is available free of charge on the ACS Publication website at:

Differentiating $X^0$ and Raman emission; Raman peak assignments of monolayer WSe$_2$ on SiO$_2$; Data from additional WSe$_2$-hBN heterostructures; Coupling between WSe$_2$ excitons and hBN phonons (PDF)

AUTHOR INFORMATION

**Corresponding Authors**

*E-mail: wangyao@hku.hk.

*E-mail: xuxd@uw.edu.

## Author Contributions

XX and WY conceived and supervised the experiments. CMC fabricated the sample and performed the measurements assisted by AMJ. CMC, XX, HY, WY analyzed the data. JY and DGM provided and characterized the bulk WSe$_2$. TT and KW provided BN crystals. CMC, XX, WY, HY wrote the paper. All authors discussed the results.

## Notes

The author declare no competing financial interest.

ACKNOWLEDGMENT

The authors acknowledge Robert Merlin for helpful discussion. This work is mainly supported by the Department of Energy, Basic Energy Sciences, Materials Sciences and Engineering Division (DE-SC0008145 and SC0012509). H.Y. and W.Y. are supported by the Croucher Foundation (Croucher Innovation Award), and the RGC and UGC of Hong Kong (HKU17305914P, HKU9/CRF/13G, AoE/P-04/08). J.Y. and D.G.M. are supported by US DoE, BES, Materials Sciences and Engineering Division. XX acknowledges a Cottrell Scholar Award, support from the State of Washington funded Clean Energy Institute, and Boeing Distinguished Professorship.

REFERENCES

(1) Chen, J.-H.; Jang, C.; Xiao, S.; Ishigami, M.; Fuhrer, M. S. Intrinsic and Extrinsic Performance Limits of Graphene Devices on SiO$_2$. *Nat. Nanotechnol.* **2008**, *3*, 206–209.
(2) Dean, C. R.; Young, A. F.; Meric, I.; Lee, C.; Wang, L.; Sorgenfrei, S.; Watanabe, K.; Taniguchi, T.; Kim, P.; Shepard, K. L.; Hone, J. Boron Nitride Substrates for High-Quality Graphene Electronics. *Nat. Nanotechnol.* **2010**, *5*, 722–726.
(3) Lee, J. J.; Schmitt, F. T.; Moore, R. G.; Johnston, S.; Cui, Y.-T.; Li, W.; Yi, M.; Liu, Z. K.; Hashimoto, M.; Zhang, Y.; Lu, D. H.; Devereaux, T. P.; Lee, D.-H.; Shen, Z.-X. Interfacial Mode Coupling as the Origin of the Enhancement of T$_c$ in FeSe Films on SrTiO$_3$. *Nature* **2014**, *515*, 245–248.
(4) Hunt, B.; Sanchez-Yamagishi, J. D.; Young, A. F.; Yankowitz, M.; LeRoy, B. J.; Watanabe, K.; Taniguchi, T.; Moon, P.; Koshino, M.; Jarillo-Herrero, P.; Ashoori, R. C. Massive Dirac Fermions and Hofstadter Butterfly in a van Der Waals Heterostructure. *Science* **2013**, *340*, 1427–1430.
(5) Moody, G.; Kavir Dass, C.; Hao, K.; Chen, C.-H.; Li, L.-J.; Singh, A.; Tran, K.; Clark, G.; Xu, X.; Berghäuser, G.; Malic, E.; Knorr, A.; Li, X.; Mak, K. F.; Lee, C.; Hone, J.; Shan, J.; Heinz, T. F.; Splendiani, A. *et al.* Intrinsic Homogeneous Linewidth and Broadening Mechanisms of Excitons in Monolayer Transition Metal Dichalcogenides. *Nat. Commun.* **2015**, *6*, 8315.
(6) Rivera, P.; Seyler, K. L.; Yu, H.; Schaibley, J. R.; Yan, J.; Mandrus, D. G.; Yao, W.; Xu, X. Valley-Polarized Exciton Dynamics in a 2D Semiconductor Heterostructure. *Science* **2016**, *351*, 688–691.
(7) Hamdi, I.; Meskini, N. Ab Initio Study of the Structural, Elastic, Vibrational and Thermodynamic Properties of the Hexagonal Boron Nitride: Performance of LDA and

GGA. *Phys. B Condens. Matter* **2010**, *405*, 2785–2794.

(8) Serrano, J.; Bosak, A.; Arenal, R.; Krisch, M.; Watanabe, K.; Taniguchi, T.; Kanda, H.; Rubio, A.; Wirtz, L. Vibrational Properties of Hexagonal Boron Nitride: Inelastic X-Ray Scattering and *Ab Initio* Calculations. *Phys. Rev. Lett.* **2007**, *98*, 095503.

(9) Ferrari, A. C.; Meyer, J. C.; Scardaci, V.; Casiraghi, C.; Lazzeri, M.; Mauri, F.; Piscanec, S.; Jiang, D.; Novoselov, K. S.; Roth, S.; Geim, A. K. Raman Spectrum of Graphene and Graphene Layers. *Phys. Rev. Lett.* **2006**, *97*, 187401.

(10) Narula, R.; Reich, S. Double Resonant Raman Spectra in Graphene and Graphite: A Two-Dimensional Explanation of the Raman Amplitude. *Phys. Rev. B* **2008**, *78*, 165422.

(11) Sun, L.; Yan, J.; Zhan, D.; Liu, L.; Hu, H.; Li, H.; Tay, B. K.; Kuo, J.-L.; Huang, C.-C.; Hewak, D. W.; Lee, P. S.; Shen, Z. X. Spin-Orbit Splitting in Single-Layer $MoS_2$ Revealed by Triply Resonant Raman Scattering. *Phys. Rev. Lett.* **2013**, *111*, 126801.

(12) del Corro, E.; Terrones, H.; Elias, A.; Fantini, C.; Feng, S.; Nguyen, M. A.; Mallouk, T. E.; Terrones, M.; Pimenta, M. A. Excited Excitonic States in 1L, 2L, 3L, and Bulk $WSe_2$ Observed by Resonant Raman Spectroscopy. *ACS Nano* **2014**, *8*, 9629–9635.

(13) Goldstein, T.; Chen, S.-Y.; Tong, J.; Xiao, D.; Ramasubramaniam, A.; Yan, J.; Mak, K. F.; Xu, X.; Yao, W.; Xiao, D.; Heinz, T. F.; Xiao, D.; Liu, G.; W. Bin, F.; Xu, X.; Yao, W.; Zhang, L.; Niu, Q.; Jariwala, D. *et al.* Raman Scattering and Anomalous Stokes–anti-Stokes Ratio in $MoTe_2$ Atomic Layers. *Sci. Rep.* **2016**, *6*, 28024.

(14) Jones, A. M.; Yu, H.; Schaibley, J. R.; Yan, J.; Mandrus, D. G.; Taniguchi, T.; Watanabe, K.; Dery, H.; Yao, W.; Xu, X. Excitonic Luminescence Upconversion in a Two-Dimensional Semiconductor. *Nat Phys* **2016**, *12*, 323–327.

(15) Chow, C. M.; Yu, H.; Jones, A. M.; Schaibley, J. R.; Koehler, M.; Mandrus, D. G.; Merlin, R.; Yao, W.; Xu, X. Phonon-Assisted Oscillatory Exciton Dynamics in Monolayer $MoSe_2$. arXiv Preprint. *arXiv*: 1701.02770v1, **2017**.

(16) He, K.; Kumar, N.; Zhao, L.; Wang, Z.; Mak, K. F.; Zhao, H.; Shan, J. Tightly Bound Excitons in Monolayer $WSe_2$. *Phys. Rev. Lett.* **2014**, *113*, 026803.

(17) Geick, R.; Perry, C. H.; Rupprecht, G. Normal Modes in Hexagonal Boron Nitride. *Phys. Rev.* **1966**, *146*, 543–547.

(18) Jin, C.; Kim, J.; Suh, J.; Shi, Z.; Chen, B.; Fan, X.; Kam, M.; Watanabe, K.; Taniguchi, T.; Tongay, S.; Zettl, A.; Wu, J.; Wang, F. Interlayer Electron–phonon Coupling in $WSe_2$/hBN Heterostructures. *Nat. Phys.* **2016**, Advance Online Publication, http://dx.doi.org/10.1038/nphys3928.

(19) Wang, G.; Glazov, M. M.; Robert, C.; Amand, T.; Marie, X.; Urbaszek, B. Double Resonant Raman Scattering and Valley Coherence Generation in Monolayer $WSe_2$. *Phys. Rev. Lett.* **2015**, *115*, 117401.

(20) Wang, G.; Marie, X.; Gerber, I.; Amand, T.; Lagarde, D.; Bouet, L.; Vidal, M.; Balocchi, A.; Urbaszek, B. Giant Enhancement of the Optical Second-Harmonic Emission of $WSe_2$ Monolayers by Laser Excitation at Exciton Resonances. *Phys. Rev. Lett.* **2015**, *114*, 097403.

(21) Poellmann, C.; Steinleitner, P.; Leierseder, U.; Nagler, P.; Plechinger, G.; Porer, M.; Bratschitsch, R.; Schüller, C.; Korn, T.; Huber, R. Resonant Internal Quantum Transitions and Femtosecond Radiative Decay of Excitons in Monolayer $WSe_2$. *Nat. Mater.* **2015**, *14*, 889–893.

(22) Chen, D.-Z. A.; Chen, G. Measurement of Silicon Dioxide Surface Phonon-Polariton Propagation Length by Attenuated Total Reflection. *Appl. Phys. Lett.* **2007**, *91*, 121906.

(23) Zhang, L. M.; Andreev, G. O.; Fei, Z.; McLeod, A. S.; Dominguez, G.; Thiemens, M.; Castro-Neto, A. H.; Basov, D. N.; Fogler, M. M. Near-Field Spectroscopy of Silicon Dioxide Thin Films. *Phys. Rev. B* **2012**, *85*, 075419.

(24) Cataliotti, R. Phonon Spectrum and Phonon Interactions in Corundum. *J. Phys. C Solid State Phys.* **1974**, *7*, 3467–3472.

(25) Kadleíková, M.; Breza, J.; Veselý, M. Raman Spectra of Synthetic Sapphire. *Microelectronics J.* **2001**, *32*, 955–958.
TOC GRAPHIC

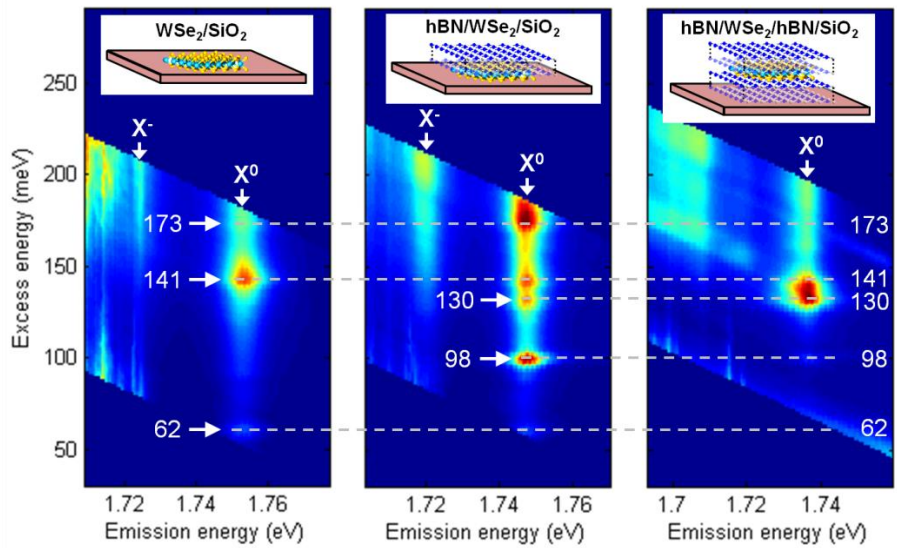

# Unusual Exciton-Phonon Coupling at van der Waals Engineered Interfaces: Supporting Information

Colin M. Chow[†], Hongyi Yu[‡], Aaron M. Jones[†], Jiaqiang Yan[§], David G. Mandrus[§,∥], Takashi Taniguchi[⊥], Kenji Watanabe[⊥], Wang Yao[‡], and Xiaodong Xu[†,#,*]

[†] Department of Physics, University of Washington, Seattle, Washington 98195, USA.
[‡] Department of Physics and Centre of Theoretical and Computational Physics, University of Hong Kong, Hong Kong, China.
[§] Materials Science and Technology Division, Oak Ridge National Laboratory, Oak Ridge, Tennessee 37831, USA.
[∥] Department of Materials Science and Engineering, University of Tennessee, Knoxville, Tennessee 37996, USA.
[⊥] Advanced Materials Laboratory, National Institute for Materials Science, Tsukuba, Ibaraki 305-0044, Japan.
[#] Department of Materials Science and Engineering, University of Washington, Seattle, Washington 98195, USA.
*E-mail: xuxd@uw.edu

This supporting document is comprised of four sections: (i) Differentiating exciton emission and Raman scattering, (ii) Raman peak assignments of monolayer $WSe_2$ on $SiO_2$, (iii) Data from additional $WSe_2$-hBN heterostructures, and (iv) Coupling between $WSe_2$ excitons and hBN phonons. Labels for figures and tables in the supplementary materials begin with the letter "S" in order to distinguish from those in the main text.

## I. Differentiating exciton emission and Raman scattering

In the main text, we mentioned that the Raman signals in Figure 1b,c remain visible far away from the $X^0$ resonance at 1.747 eV. This is evident when the PLE spectra are plotted in log scale. An example is given in Figure S1a for an hBN-covered $WSe_2$ (hBN/$WSe_2$/$SiO_2$) sample structure. Figure S1b shows line traces of the data enclosed by the red dashes in Figure S1a, now plotted in linear scale. Here, three prominent features consisting of the $X^0$ and $X^-$ luminescence, and the Stokes-shifted hBN $A_{2u}$ Raman peaks are marked by dashed lines. The luminescence and Raman signal differ in their characteristic linewidths, of which the full-width-half-maximum (FWHM) of $X^0$ and $X^-$ luminescence is around 8 meV, while that of Raman scattering is about 5.5 meV, determined by our laser linewidth. The mechanisms of these light scattering processes are compared in Figure S1c for the regime in which excitation is far from the $X^0$ + hBN $A_{2u}$ phonon resonance. In Raman scattering, the Stokes-shifted emission lies below the excitation energy by one hBN $A_{2u}$ phonon, and shifts in parallel with the excitation laser detuning. The excitonic luminescence, however, occurs at a constant, excitation-independent energy.

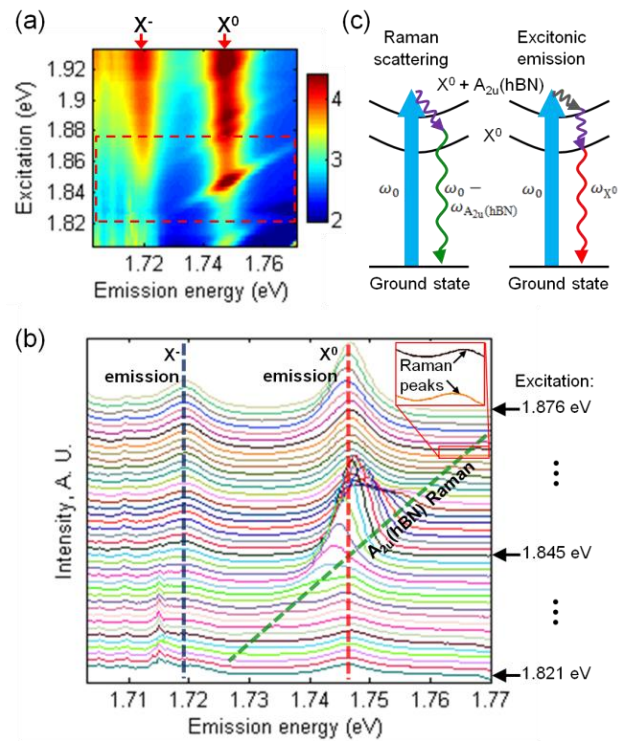

**Figure S1.** (a) PLE intensity map from data presented in Figure 1c (hBN/$WSe_2$/$SiO_2$ geometry), with log color scale. Color bar: $\log_{10}$ of PL counts. (b) Line traces of the PLE data (offset for clarity) enclosed by the red dashes in (a). Blue, red and green dashed lines mark the peak positions of $X^-$ and $X^0$ emission, and hBN $A_{2u}$ Raman signal, respectively. Inset: zoomed in view of the boxed region, showing Raman peaks away from the $X^0$ resonance. (c) Schematic comparison between Raman scattering and excitonic emission. Blue arrows represent excitation at frequency $\omega_0$, while wavy arrows denote phonon emission (purple and grey), Stokes-shifted Raman (green) and $X^0$ emission (red).

In Figure S1b, both $X^0$ and $X^-$ emission peaks show a general trend of becoming weaker as the excitation laser energy is decreased. The Raman signal, however, becomes strongly enhanced around an excitation of 1.845 eV, where the Stokes peak is resonant with the $X^0$ transition. Here, a double-resonant Raman configuration is formed, as illustrated in Figure 1d in the main text. The signal at the $X^0$ resonance now consists of both $X^0$ emission and Raman scattered light. To estimate the enhancement of the Raman signal, we assume that the $X^0$ emission is largely unaffected by the resonant excitation of the $X^0$ + hBN $A_{2u}$ phonon mode. This is justified by the fact that the linewidths of both signals remain distinct, with the Raman peak width mirroring that of the excitation laser. After removing the broader $X^0$ peak, we find that the intensity of the Raman signal is enhanced 90 times relative to off-resonant excitation. We therefore conclude that interfacial exciton-phonon coupling between monolayer $WSe_2$ and atomically-thin hBN provides nearly two orders of magnitude amplification of the Raman signal.

## II. Raman Peak Assignments of Monolayer $WSe_2$ on $SiO_2$

Vibrational modes in monolayer $WSe_2$ and their associated Raman spectra have been extensively studied[1,2]. Our Raman spectra closely match those reported in the literature. Figure S2a–c show Raman peak positions for monolayer $WSe_2$ on $SiO_2$ substrate, with their mode assignments given in Table S1.

**Table S1. Assignment of Raman peaks indicated in Figure S2a–c according to ref. 2**

| Raman peak (meV) | Assignment[2] |
|---|---|
| 12.0 | $E'(M)^{TO} - LA(M)$ |
| 14.6 | $E'(M)^{LO} - LA(M)$ |
| 17.0 | $A'(M) - LA(M)$ |
| 20.0 | $E'(M)^{LO} - TA(M)$ |
| 27.3 | $LA(M) + TA(M)$ |
| 29.6 | $LA(M) + ZA(M)$ |
| 31.0 | $A'_1(\Gamma)$ or $E'(\Gamma)$ |
| 32.2 | $2\,LA(M)$ |
| 43.8 | $E''$ |
| 46.4 | $E'(M)^{LO} + LA(M)$ |
| 49.4 | $A'_1(M) + LA(M)$ |
| 62.0 | $2\,A'_1(\Gamma)$ |

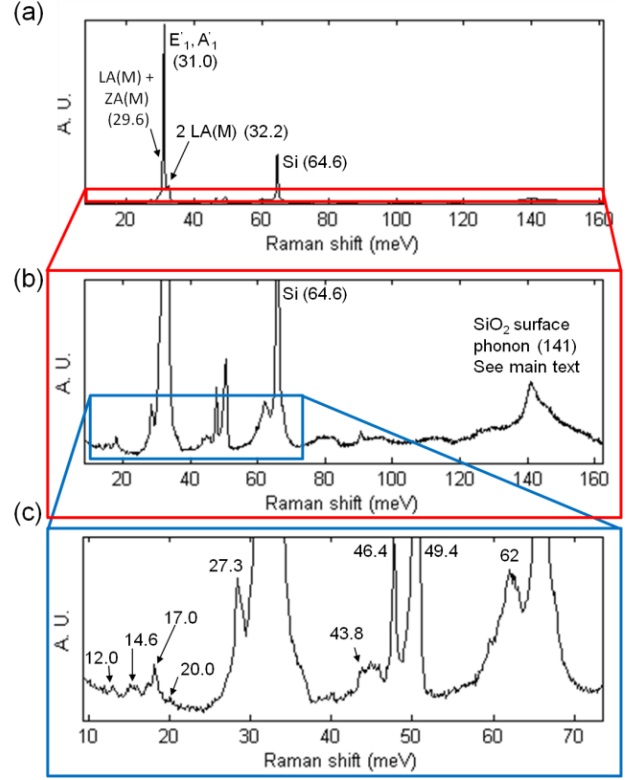

**Figure S2.** (a) Similar Raman spectrum as presented in Figure 2f in the main text, with dominant peaks labeled. (b) Magnified view of the region enclosed in the red box in (a). (c) Magnified view of the region enclosed in the blue box in (b). Raman peak positions are indicated in units of meV.

## III. Data from additional $WSe_2$-hBN heterostructures

All features described in the main text have been reproduced in a second set of samples, as shown in Figure S3. Additionally, we performed PLE spectroscopy of a $WSe_2$ monolayer placed on an atomically-thin (~5 nm) sheet of hBN ($WSe_2$/hBN/$SiO_2$ geometry). The results shown in Figure S3c,g are qualitatively similar to those obtained from hBN-encapsulated $WSe_2$ (hBN/$WSe_2$/$SiO_2$) samples. An important difference, as discussed in the main text, is the weaker 141-meV emission relative to the 130-meV peak. This is due to screening of the $SiO_2$ surface phonon coupling, the source of the 141-meV signal, by the interposing hBN sheet.

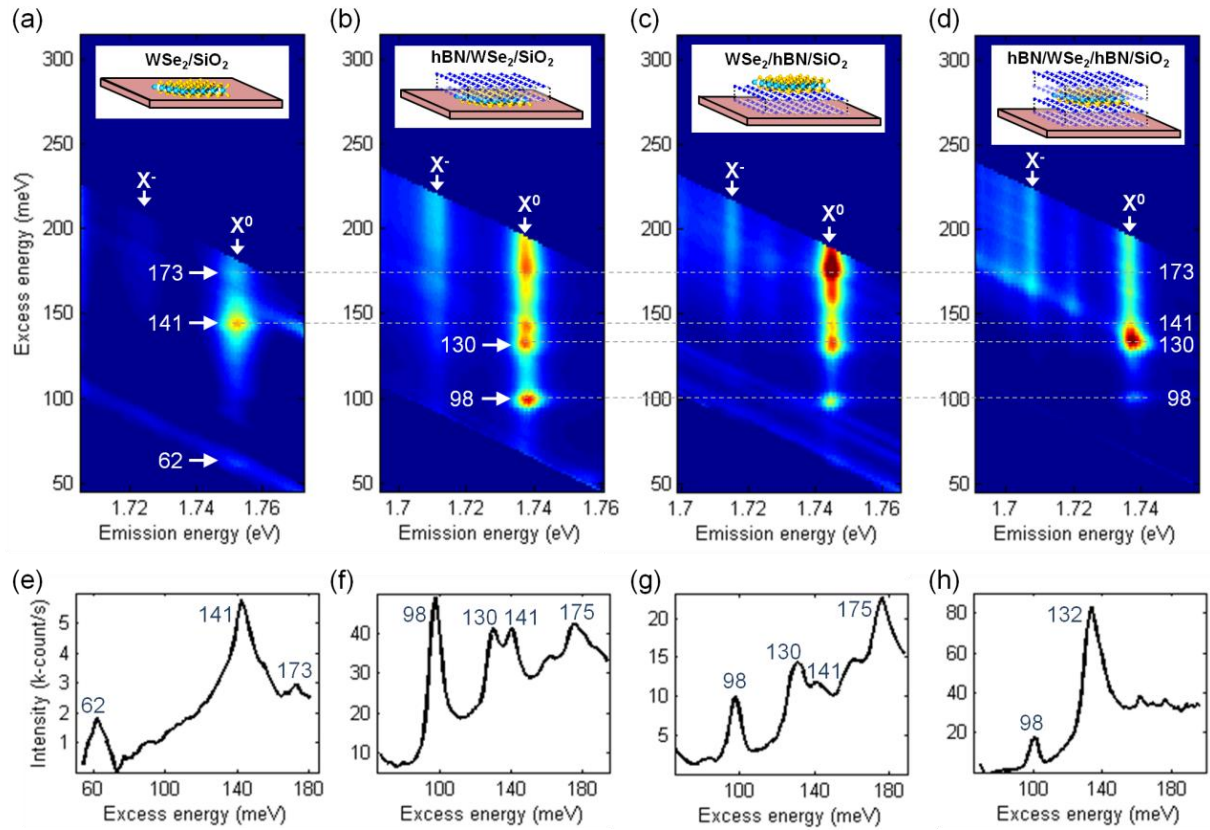

**Figure S3.** (a–d) PLE intensity spectra from a second set of samples, plotted with the excess energy as the vertical axis, for sample structures: WSe$_2$/SiO$_2$, hBN/WSe$_2$/SiO$_2$, WSe$_2$/hBN/SiO$_2$ and hBN/WSe$_2$/hBN/SiO$_2$, respectively, as illustrated in corresponding insets. (e–h) Vertical line cuts at the X$^0$ resonance from PLE maps (a–d), respectively.

## IV. Coupling between WSe$_2$ excitons and hBN phonons

The 98-meV mode in multilayer hBN corresponds to the out-of-plane optical phonons (ZO) A$_{2u}$ or B$_{1g}$, which are Raman inactive due to the D$_{6h}$ symmetry of the multilayer hBN. Forming WSe$_2$/hBN heterostructures lowers the symmetry, and the hBN out-of-plane phonons can in principle appear in Raman spectrum. Moreover, through direct or indirect coupling to the WSe$_2$ exciton, the hBN out-of-plane phonons can also become visible in PLE or Raman spectroscopy. In the following, we explore the physical mechanism behind the interaction between hBN out-of-plane phonons and the WSe$_2$ exciton.

Since the B(N) atoms carry positive(negative) charges, the relative displacement, $\delta z'$, between B and N atoms corresponds to an out-of-plane electric dipole, $\delta D'$ (see Figure S4a). Similarly, the tri-atomic planes in the WSe$_2$ monolayer can also be viewed as two layers of electric dipoles, $\pm D$, pointing out of plane in opposite directions, and their different distances from the hBN leads to a net coupling, $V(\mathbf{r}, z)$, to the dipole $\delta D'$ associated with the hBN phonons. The WSe$_2$ ZA phonon mode changes the vertical distance between $D$ and $\delta D'$ from $z$ to $z + \delta z$ (Figure S4b), while the A$_1'$ mode changes the magnitude of $D$ to $D + \delta D$ (Figure S4c). Both phonon modes can result in a coupling between hBN and WSe$_2$ out-of-plane phonons.

With the above interlayer phonon-phonon coupling, the interaction between hBN out-of-plane phonons and WSe$_2$ excitons can be mediated by WSe$_2$ ZA or A$_1'$ phonons. In WSe$_2$, the exciton-phonon interaction matrix elements are dictated by the symmetry of the vibrational modes. As the out-of-plane mirror ($\hat{\sigma}_h$) symmetry is present in monolayer WSe$_2$, the ZA phonon has odd parity, $\hat{\sigma}_h = -1$, while the exciton has even parity, $\hat{\sigma}_h = +1$, Therefore, the first order exciton-ZA phonon coupling is forbidden. Notably, in layered materials without $\hat{\sigma}_h$ symmetry, it has recently been shown that the ZA phonon can greatly reduce carrier mobility[3,4].

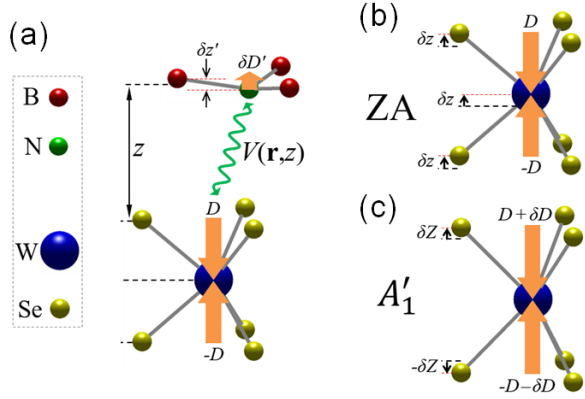

**Figure S4.** (a) Schematic illustration of phonon-induced dipole-dipole interaction between hBN and $WSe_2$. The hBN ZO phonon corresponds to a vertical relative displacement, $\delta z'$, between B and N atomic planes, thus creating an electric dipole, $\delta D'$. The $WSe_2$ atomic configuration can be viewed as two static electric dipoles $\pm D$. $V(\mathbf{r}, z)$ is the interaction between $\delta D'$ and $D$. (b) The $WSe_2$ ZA phonon mode corresponds to an overall vertical displacement, $\delta z$, of all W and Se atoms. (c) The $WSe_2$ $A_1'$ phonon mode corresponds to a change in the magnitudes of the two static electric dipoles: $\pm D \rightarrow \pm(D + \delta D)$.

In hBN-covered $WSe_2$ samples (hBN/$WSe_2$/$SiO_2$ geometry), the one-sided contact with hBN breaks the $\hat{\sigma}_h$ symmetry in the $WSe_2$, thus an exciton is now allowed to absorb/emit a ZA phonon, which can then couple to the hBN out-of-plane phonons, leading to the appearance of a strong 98-meV PLE/Raman peak. On the other hand, $\hat{\sigma}_h$ symmetry is preserved (or weakly broken) in hBN-sandwiched $WSe_2$ samples (hBN/$WSe_2$/hBN/$SiO_2$ geometry), which then suppresses the coupling of the $WSe_2$ exciton to the ZA phonon. However, second-order exciton-phonon coupling involving two ZA phonons is still allowed by the $\hat{\sigma}_h$ symmetry. As a result, a weak 98 meV PLE/Raman peak is observed in hBN-sandwiched $WSe_2$ samples.

The 130-meV PLE and Raman peaks involve the emission of an $A_1'$ phonon and an hBN out-of-plane phonon. This can be realized through the emission of two $WSe_2$ $A_1'$ phonons by the exciton, followed by the coupling of one $WSe_2$ $A_1'$ phonon to the hBN phonon. Note that the $A_1'$ phonon has $\hat{\sigma}_h = +1$, rendering its coupling to the exciton insensitive to the $\hat{\sigma}_h$ symmetry of the sample stack. Beyond the 130 meV phonon resonance, a laser with an energy of 130 meV above the 1s exciton ground state also lies near the 2s exciton state resonance, yielding a double-resonance enhancement. This is consistent with the observation that the 130 meV PLE/Raman peak is strong in both hBN-covered and hBN-sandwiched $WSe_2$ samples.

## REFERENCES

(1) Terrones, H.; Corro, E. Del; Feng, S.; Poumirol, J. M.; Rhodes, D.; Smirnov, D.; Pradhan, N. R.; Lin, Z.; Nguyen, M. A. T.; Elías, A. L.; Mallouk, T. E.; Balicas, L.; Pimenta, M. A.; Terrones, M.; Hwang, W. S.; Wang, Q. A.; Kalantar-Zadeh, K.; Kis, A.; Coleman, J. N. *et al. Sci. Rep.* **2014**, *4*, 699–712.

(2) Zhang, X.; Qiao, X.-F.; Shi, W.; Wu, J.-B.; Jiang, D.-S.; Tan, P.-H.; Novoselov, K. S.; Neto, A. H. C.; Bonaccorso, F.; Lombardo, A.; Hasan, T.; Sun, Z.; Colombo, L.; Ferrari, A. C.; Coleman, J. N.; Lotya, M.; O'Neill, A.; Bergin, S. D.; King, P. J. *et al. Chem. Soc. Rev.* **2015**, *44*, 2757–2785.

(3) Fischetti, M. V.; Vandenberghe, W. G. *Phys. Rev. B* **2016**, *93*, 155413.

(4) Gunst, T.; Kaasbjerg, K.; Brandbyge, M. arXiv Preprint. *arXiv*: 1609.05852v2, **2016**.